\documentclass[a4paper]{article}
\usepackage[english]{babel}

\usepackage[a4paper,top=2cm,bottom=2cm,left=3cm,right=3cm,marginparwidth=1.75cm]{geometry}
\usepackage{comment}
\usepackage{amsmath}
\usepackage{mathtools}
\usepackage[makeroom]{cancel}
\usepackage{graphicx}
\usepackage[colorlinks=true, allcolors=blue]{hyperref}
\usepackage{tikzfeynman}
\usepackage{authblk}
\usepackage{cleveref}
\usepackage{todonotes}
\usepackage{amsfonts} 
\usepackage[normalem]{ulem}

\newcommand{\ep}{\ensuremath{\varepsilon}}

\begin{document}

\title{Hyperlogarithms in the theory of turbulence of infinite dimension }

\author[1,2]{Loran Ts. Adzhemyan}
\affil[1]{Saint Petersburg State University, 7/9 Universitetskaya nab., St. Petersburg, 199034, Russian Federation} 
\affil[2]{Bogolyubov Laboratory of Theoretical Physics, Joint Institute for Nuclear Research, 141980 Dubna, Russian Federation}
\author[1]{Daniil A. Evdokimov}
\author[1,2]{Mikhail V. Kompaniets \footnote{m.kompaniets@spbu.ru}}

\maketitle

\begin{abstract}
Parametric integration with hyperlogarithms so far has been successfully used in problems of high energy physics (HEP) and critical statics. In this work, for the first time, it is applied to a problem of critical dynamics, namely, a stochastic model of developed turbulence in high-dimensional spaces, which has a propagator that is non-standard with respect to the HEP: $(-i \omega + \nu k^2)^{-1}$. Adaptation of the hyperlogarithm method is carried out by choosing a proper renormalization scheme and considering an effective dimension of the space. Analytical calculation of the renormalization group functions is performed up to the fourth order of the perturbation theory, $\varepsilon$-expansion of the critical exponent $\omega$ responsible for the infrared stability of the fixed point is obtained.

\end{abstract}

\section{Introduction}

The renormalization group and the $\ep$-expansion are successfully applied to the theory of critical phenomena, both in problems of statics and stochastic dynamics \cite{V04, Zinn_book}. 
Recently, effective methods for calculating perturbation theory diagrams have been developed, which have made it possible to obtain analytical results in high orders of perturbation theory both in high energy physics \cite{GGP_5l, Lee5l, Laporta2017} and critical statics \cite{KP17, Schnetz7l, Bednyakov2021, Bednyakov:2022}. 
In the case of critical dynamics, results are more modest: analytical answers were obtained in almost all models only in one loop \cite{Forster_Nelson77},
 the second order was achieved in series of models numerically \cite{Folk_modelC, Adzhemyan1999Hmodel, Dominicis_Hmodel}, only a few were calculated in three loop order \cite{AV84, perc2023}  
and only model A~\cite{AV84} has been calculated analytically. Further calculations of the model A were performed numerically, which allowed to advance into the fourth \cite{ANS08} and fifth \cite{AEHIKKZ22, AEHIKKZ22_2} orders  recently.
The difficulties of calculations in the critical dynamics are related to the dissipative nature of the equations of motion, resulting in the violation of Lorentz invariance in difference to high energy physics models.

The computations in dynamic models with vector order parameter are even more challenging. Thus, in the theory of fully developed turbulence, the analytical answer was calculated only in one-loop approximation \cite{DM79}, and the numerical result was obtained in two-loop computation~\cite{AAKV03,AHKV05_2l,AHKS05_Prandtl2l,Prandtl2l_magn}. Significant progress was achieved by studying the asymptotic region of high-dimensional spaces, where the three-loop analytical computation was carried out in~\cite{AAGKK08}.

The hyperlogarithm method used in high energy physics is a powerful tool for analytical multiloop calculations. It was proposed by Francis Brown in the series of works \cite{Brown08, Brown09, Brown09_arx} and then further developed by Erik Panzer, resulting in a Maple implementation \textit{HyperInt} \cite{Panzer14_main}, which we apply to perform symbolic calculations of the diagrams. For a consistent introduction to this approach, we highly recommend a detailed work \cite{Panzer_thesis}.

Fundamentally, the hyperlogarithm method is based on the well-known fact that some Feynman integrals can be expressed in terms of rational combinations of multiple zeta values, which are a special case of multiple polylogarithms. The hyperlogarithms are just alternatively written multiple polylogarithms, defined as iterated integrals. They possess some useful properties and relations which make them suitable for performing symbolic integration for a broad class of Feynman diagrams \cite{Panzer14_main}. The hyperlogarithm method has been successfully applied in a number of works, e.g., the 6-\cite{KP17} and 7-loop~\cite{Schnetz23} studies of the static $\phi^4$ model  (in the latter, it was used along with the graphical function approach, which was the main calculation technique), and the 4-loop study of QCD \cite{AMPS16, AMPS20, AMPS21}. 

{
However, not every Feynman integral can be assessed with this method. First of all, the critical dimension of a theory (the dimension of the space at $\varepsilon=0$) generating these integrals must be even, which is the case in the above-mentioned models. In the turbulence theory, this requirement is of a different nature. Namely, the propagators depend on the dimensionality of the space, whereas the parameter $\varepsilon$ does not depend on it. Fortunately, asymptotic expressions for diagrams in high-dimensional spaces can be interpreted as integrals in a space with effective dimension $D=2-2\varepsilon$, even at $\varepsilon=0$.}
{
Another significant limitation is the requirement for linear reducibility, a criterion that allows an integrand to be expressed through hyperlogarithms at each step of integration and expressed through rational linear combinations of multiple zeta values.
}
In some cases, this property can be restored with an appropriate choice of order of integration {
or proper change of variables}, but often the hyperlogarithm method needs to be accompanied by other computation techniques.  

The hyperlogarithm method is applicable only for convergent massless integrals, which are free of subdivergences. Unlike linear reducibility, this requirement can be satisfied with a proper choice of a renormalization scheme that will reduce evaluating Z-factors (renormalization constants) and renormalization group (RG) functions ($\beta$-function and anomalous dimension) to the calculation of finite integrals. 

In this work, we adapt a hyperlogarithm parametric integration method for calculations in the stochastic model of developed turbulence in high-dimensional spaces which is the limit of the regular developed turbulence at  $d\to\infty$. 
The consideration of this asymptotics in this case is interesting because in the region $d\to\infty$ the anomalous exponents presumably vanish and Kolmogorov’s theory (K41) becomes valid. 
In this model, we obtain the 4-loop analytical estimation for a critical exponent $\omega$, which governs the stability of a renormalization group fixed point. 

As mentioned above, the dimension of the theory must be even. In our theory we can consider the integrals with effective space dimension $D=2-2\ep$. 
{
The aforementioned requirement of linear reducibility is satisfied without the need for additional manipulations for all integrals at least up to 4 loops.}
Lastly, we require a proper renormalization scheme, to avoid subdivergences. For this purpose, we use a scheme proposed in \cite{AKNS13} in which the RG functions are directly expressed in terms of renormalized massless integrals.
 
It is worth mentioning that the considered model is relatively simple in comparison with the others due to the absence of vertex divergences. For example, in the 6-loop study of $\phi^4$ \cite{KP17}, where the hyperlogarithm method was the main calculation tool, the authors had to combine two renormalization schemes: a modified BPHZ-like approach \cite{Brown_Kreimer13j} to construct convergent integrals, which are suitable for parametric integration, and the minimal subtraction (MS) scheme to relate the calculated diagrams to the renormalization constants. In our case, the application of the MS scheme is not necessary, and the RG functions are calculated directly without evaluating the renormalization constants.  However, there is a noticeable complexity compared to the $\phi^4$ model. In $\phi^4$, in the massless scheme, subgraphs with two external legs have a trivial dependence on an external momentum.
It allows to factor them from the diagrams and evaluate separately, which significantly simplifies the calculations. In the model considered, such procedure is impossible due to the presence of not only an external momentum, but also an external frequency.

To explain the algorithm automated in \textit{HyperInt}, in Section \ref{HL_sec} we briefly discuss the main concepts behind the application of the hyperlogarithms to the parametric multiple integration. We do not aim to provide a comprehensive introduction to this broad topic, but rather focus on illustrating the appearance of zeta values within the framework of hyperlogarithm parametric integration. As a demonstration, we include the integration of a 3-loop diagram. The interested reader may refer to \cite{Brown08, Golz13_thesis}\footnote{In the thesis \cite{Golz13_thesis}, we spot a minor misprint in the formula (4.65): the wrong sign of $\zeta_2$.}, where the authors provide more illustrative examples of the integration process on two loop t-bubble (Chetyrkin) diagram.

Finally, the dynamic nature of this model has not been covered in the previous applications of the hyperlogarithm method to static models and requires special consideration. 
This is mainly related to the proper renormalization of the dynamic diagrams within the chosen scheme, which is performed using the Bogoliubov-Parasiuk $R$-operation. We provide an efficient automated implementation that allows us to construct convergent integrals for all diagrams up to four loops. All other computation steps, including generation, selection, and integration of the diagrams, are automated as well.

\section{Model}

 The stochastic model of the fully developed, isotropic, homogeneous turbulence of the incompressible fluid is described by the stochastic Navier-Stokes equation with a random stirring force \cite{V04,AAVBook}:
\begin{equation}
\partial_t v_i = - \partial_i P - (v_k \partial_k) v_i + \nu_0 \partial^2 v_i + f_i \;,
\label{ns}
\end{equation}
where $v_i$ is the velocity field, $P$ is the pressure, $\nu_o$ is the kinematic viscosity, and $f_i$ is the random force. Equation \eqref{ns} is complemented with the incompressibility condition $\partial_i v_i =0 $ that leads to the transverness of the velocity field. For the random force, the Gaussian statistics with zero mean and pair correlator of form \eqref{df} is assumed: 

\begin{align}
\left< f_i(\vec{x}_1, t_1) f_j( \vec{x}_2,t_2)\right> \equiv  D^f_{ij}(\vec{x}_1-\vec{x}_2,t_1-t_2)\nonumber\\
D^f_{ij}(\vec{k},t)=\delta(t)P_{ij}(\vec{k}) d_{f}({k})\;,\qquad P_{ij}(\vec{k})=\delta_{ij}-\frac{k_i k_j}{k^2} \; .
\label{df}
\end{align}
 The presence of the transverse projector $P_{ij}(\mathbf{k})$ in \eqref{df} is a consequence of the incompressibility condition. The function $d_{f}(\mathbf{k})$  describes injection of energy into the system. In the inertial interval of the wavenumbers $m\ll k \ll k_{dis}$ ($m^{-1}=L$  is external turbulence scale, $k_{dis}$ is dissipative scale), one can use the power-law model
\begin{equation}
d_{f}({k})=D_0 k^{4-d-2\ep} F\left(\frac{m}{k}\right),
\label{df2}
\end{equation}
 where $\varepsilon$ plays the same role as the $4-d$ parameter in the Wilson theory of the phase transitions. Its physical value is equal to $2$, which corresponds to the ideal pumping of energy with infinitely large eddies.

  The stochastic model \eqref{ns} is equivalent to the quantum-field action with a doubled set of fields~\cite{V04,AAVBook}:
\begin{equation}
S_0=\frac{1}{2}v'D^fv' + v' \left[ -\partial_t v - (v\partial)v + \nu_0 \partial^2 v\right].
\label{s0}
\end{equation}
Above, all necessary integrations in coordinates and times and summations over indices are implied. The contribution of the pressure in \eqref{s0} is omitted due to the transverseness  of the auxiliary field $v'$.

The diagrams in the perturbation theory with action \eqref{s0} contain ultraviolet (UV) divergences in the limit $\ep \to +0$ , which take place only in the 1-irreducible correlation function $\Gamma_{ij}^{(0)}=\left< v_i v_j'\right>_{1-irr}^{(0)}$. The Galilean invariance of the stochastic Navier-Stokes equation leads to the fact that there are no divergences in the 1-irreducible correlation function $\left< v_i v_j v_k'\right>_{1-irr}^{(0)}$. To cancel divergences in function $\Gamma_{ij}^{(0)}$ the counterterm of the form $v'\partial^2 v$ is required. The renormalized action is given by

\begin{equation}
S=\frac{1}{2}v'D^fv' + v' \left[ -\partial_t v - (v\partial)v + \nu Z_\nu \partial^2 v\right],
\label{sr}
\end{equation}
 which is obtained from \eqref{s0} by the multiplicative renormalization of the parameters:
\begin{equation}
D_0=g_0 \nu^3_0 = g \mu^{2\ep}\nu^3\;, \quad  \nu_0 = \nu Z_\nu \;, \quad g_0 = g \mu^{2\ep} Z_g \;, \quad Z_g = Z_\nu^{-3}\;,
\label{renorm}
\end{equation}
 where $\mu$ is the renormalization mass and $g$ is the dimensionless renormalized charge. The renormalization of the fields is not required.

Let $\Gamma_{ij}(\vec{k},\omega)$ denote the 1-irreducible function $\left< v_i v_j'\right>_{1-irr}$ calculated using the action \eqref{sr} with $Z_\nu=1$ (corresponding to the basic action)~\cite{V04,AAVBook}. This function is proportional to the transverse projector:

\begin{equation}
\Gamma_{ij}(\vec{k},\omega) = P_{ij}(\vec{k}) \Gamma(k,\omega)\; , \quad \Gamma(k,\omega)=\frac{\Gamma_{ii}(k,\omega)}{d-1}\,.
\end{equation}
Let us define a normalized function 
\begin{equation}
\overline{\Gamma}(k,\omega) \equiv \frac{\Gamma(k,\omega)}{-\nu k^2},
\label{gch}
\end{equation}
 which is equal to one in the loopless approximation. 

In the earlier papers \cite{AAGKK09_turb_3l,AKKS17_turb} we  used massive renormalization scheme and the renormalization constant $Z_\nu$ was chosen to satisfy the following condition for {
renormalized function} $\overline{\Gamma}^R$ in the normalization point $k=0\;, \omega=0\;, m=\mu$: 
\begin{equation}
\overline{\Gamma}^R|_{k=0,\omega=0,m=\mu} = 1 \;.
\label{oldtn}
\end{equation}
To provide the infrared (IR) regularization, in \eqref{df2} the function $F(m/k)$ was chosen in the form of a sharp cutoff $F(m/k)=\theta(k-m)$, where $\theta(...)$ is the Heaviside function.
In this paper, in order to use the hyperlogarythm method, we consider the massless theory with $m=0$ and set $F(m/k)\equiv 1$. Consequently, the normalization point is changed from \eqref{oldtn} to $(\omega=0\;, k =\mu)$, so we require
\begin{equation}
\overline{\Gamma}^R|_{k=\mu,\omega=0} = 1\;.
\label{usl}
\end{equation}
In that case, the IR regularization is guaranteed by the finiteness of the momentum $k$.

The free propagators corresponding to the model \eqref{sr},\eqref{df},\eqref{df2},\eqref{renorm} in the $(k,t)$ representation take the form
\begin{equation}
\langle v_i(t_1) v_j(t_2)\rangle = \frac{g \mu^{2\ep}\nu^2}{2} k^{2-d-2\ep} e^{-E_k\cdot|t_1-t_2|} P_{ij}(\vec{k}) =\vcenter{\hbox{ 
\begin{tikzpicture}[node distance=1.5cm]
\coordinate (v1); 
\coordinate[right=of v1] (v2);
\arcl{v1}{v2}{0};
\end{tikzpicture}}}\; ,
\label{l1}
\end{equation}
    \begin{equation}
    \langle v_i(t_1) v_j'(t_2)\rangle = \theta(t_1-t_2) e^{-E_k \cdot(t_1-t_2)} P_{ij}(\vec{k}) = \vcenter{\hbox{
    \begin{tikzpicture}[node distance=1.5cm]
    \coordinate (v1); 
    \coordinate[right=of v1] (v2);
    \coordinate(e2) at ($ (v1)!.15!(v2) $);
    \coordinate[above=of e2] (sv1);
    \coordinate[below=of e2] (sv2);
    \arcl{v1}{v2}{0};
    \draw ($ (v2)!.15!0:(sv1) $) -- ($ (v2)!.15!0:(sv2) $);
    \end{tikzpicture}}}\; ,
    \label{l2}
    \end{equation}
\begin{equation}
\langle v_i'(t_1) v_j'(t_2)\rangle = 0 \;,
\label{l3}
\end{equation}
where ``energy'' $E_k=\nu k^2$.
The interaction in \eqref{sr} is represented by a triple vertex $-v'(v\partial)v = v'_m V_{m n p} v_n v_p$ with a vertex factor
\begin{equation}
V_{m n p}= i k_n \delta_{m p}= \vcenter{\hbox{
\begin{tikzpicture}[node distance=1.5cm]
    \coordinate (v1); 
    \coordinate[right=of v1] (v2);
    \coordinate(e2) at ($ (v1)!.15!(v2) $);
    \coordinate[above=of e2] (sv1);
    \coordinate[below=of e2] (sv2);
    \coordinate[right=of v2] (e3);
    \coordinate[above=of e3] (v3);
    \coordinate[below=of e3] (v4);
    \coordinate[vertexL](e4) at ($ (v2)!.15!(v3) $);
    \arclls{v1}{v2}{0}{\quad m}{below=0.15cm};
    \arclls{v3}{v2}{0}{n}{below=0.20cm};
    \arclls{v4}{v2}{0}{p}{left=0.25cm};
    \draw ($ (v2)!.15!0:(sv1) $) -- ($ (v2)!.15!0:(sv2) $);
    \end{tikzpicture}}}
    \label{vertex}
\end{equation}
where $k_n$ is the momentum argument of the field $v'$. The crossed line in \eqref{vertex} corresponds to the field $v'$, the line with the dot corresponds to the field $v_n$ contracted with $i k_n$, and the plain line represents the field $v_p$. The asymmetric form of the interaction vertex turns out to be more suitable for further analysis of the diagrams in the $d\to \infty$ limit.

We define the perturbation series for the function $\overline{\Gamma}$ \eqref{gch} as
{
\begin{equation}
\overline{\Gamma}(k,\omega,\mu) = 1 + \sum_{n\ge 1} {h}^n \mu^{2n\ep} \sum_{i} \chi_n^{(i)}(k,\omega)\;, \quad h\equiv g \frac{ S_d }{(2\pi)^d}\;,
\label{gbar}
\end{equation}
where the index $i$ runs over all $n$-loop diagrams of the function $\overline{\Gamma}$. For convenience, we introduce a normalized charge $h$, in which $S_d$ is the surface area of the $d$-dimensional unit sphere.
}

\section{Renormalization group, RG functions expressed in terms of the renormalized Green functions.}

As in the MS scheme, the renormalization constants $Z_\nu$ and $Z_g$ depend only on the dimension of the space $d$ and the parameter $\ep$ and do not depend on the renormalization mass $\mu$. The renormalization group equations can be obtained from the independence of the non-renormalized Green functions on the parameter $\mu$ with fixed $\nu_0$ and $g_0$. They are completely identical to the ones in the MS scheme. In particular, for the 1-irreducible renormalized function $\Gamma^R$, one has
\begin{equation}
(\mu\partial_\mu+\beta\partial_g-\gamma_\nu \nu\partial_\nu)\Gamma^R=0 \;,
\label{rgeq}
\end{equation}
where
\begin{equation}
\gamma_i
=\frac{-2\ep g\partial_g \ln Z_i}{1+g\partial_g \ln Z_g}, \quad \beta=-g(2\ep+\gamma_g)=-g(2\ep-3\gamma_\nu).
\label{gi}
\end{equation}
The latter equation in \eqref{gi} is a consequence of the relation between the renormalization constants $Z_g$ and $Z_\nu$ in \eqref{renorm}. The RG fixed point is determined by the condition $\beta(g_*)=0$. Its stability is governed by the correction exponent $\omega$
\begin{equation}
\label{omega_def}
\omega=\partial_g \beta(g)|_{g=g_*}\;.
\end{equation}
The fixed point is stable if $\omega>0$. The Gaussian fixed point $g_*=0$ is unstable due to $\omega=-2\ep<0$. A nontrivial fixed point is determined by the condition
\begin{equation}
\gamma_\nu(g_*)=\frac{2\ep}{3}\;.
\end{equation}
Therefore, the $\ep$-expansion of $\gamma_\nu(g_*)$  that determines the scaling exponents cuts on the first term which is the consequence of the relations \eqref{renorm}. At the value $\ep=2$, this fact leads to the exact results of the K41 theory, in particular the famous 5/3 law for the energy spectrum. As for the exponent $\omega$, the series is infinite, and the key problem is to verify the stability of the fixed point in higher loops of perturbation theory. Our main goal is to advance to the fourth order of the expansion for the exponent $\omega$.

Let us reduce the function $\gamma_\nu$ to the form that is convenient for its calculation with the hyperlogarithm method. Firstly, we need to express it in terms of finite integrals, which is done similar to $\phi^4$ theory \cite{KP17}. After that, while in \cite{KP17} the authors restored integrals within the MS scheme, we are able to write the RG functions directly.
For this purpose, it is convenient to express $\gamma_\nu$ in terms of the normalized function $\overline \Gamma ^R$, using \eqref{rgeq}:
\begin{equation}
(\mu\partial_\mu+\beta\partial_g-\gamma_\nu \nu\partial_\nu)\overline \Gamma^R=\gamma_\nu \overline \Gamma^R .
\end{equation}
Considering this equation in the normalization point $(k=\mu, \omega=0)$ and taking into account the following corollaries of \eqref{usl}
\begin{equation}
 \overline \Gamma^R|_{k=\mu,\omega=0}=1, \quad \partial_g \Gamma^R|_{k=\mu,\omega=0}=0,  \quad \partial_\nu \Gamma^R|_{k=\mu,\omega=0}=0 ,
\end{equation}
one obtains
\begin{equation}
\gamma_\nu= (\mu \partial_\mu \overline\Gamma^R)|_{k=\mu,\omega=0}\;.
\label{gnu1}
\end{equation}
Using the dependence of the dimensionless function $\overline \Gamma ^R$ on the ratio $k/\mu$, the equation \eqref{gnu1} can be rewritten as
\begin{equation}
\gamma_\nu= -(k\partial_k \overline\Gamma^R)|_{k=\mu,\omega=0}\;.
\label{gnu2}
\end{equation}

The dimensionless $\overline \Gamma^R|_{\omega=0} (k/\mu)$ does not depend on the viscosity $\nu$, allowing one to simplify \eqref{gnu2} by setting $\nu=1, \mu=1$. For that, the dimensionless momentum $\tilde k=k/\mu$ should be used, for which we keep the symbol $k$
\begin{equation}
\gamma_\nu= -(k\partial_k \overline\Gamma^R)|_{k=1,\omega=0}\;.
\label{gnu3}
\end{equation}  
For the next step of the preparation for applying the hyperlogarithm method, the obtained equation \eqref{gnu3} for the anomalous dimension needs to be expressed with Bogoliubov's $R$-operation:
\begin{equation}
\overline\Gamma^R=R\overline\Gamma=(1-K)R'\overline\Gamma\;.
\end{equation}  
Here, the $R'$-operation subtracts all UV subdivergences, and the $(1-K)$ operation eliminates the remaining superficial divergence. The $R'$-operation can be expressed as
\begin{equation}
R'\overline\Gamma=\prod_j(1-K)_j\overline\Gamma,
\label{Rzav}
\end{equation}  
where the product runs over all divergent subgraphs of the diagrams from $\overline\Gamma$, starting from the minimal one in case of nested subgraphs. It is worth mentioning that there are no overlapping subgraphs due to the absence of vertex divergences. This leads to a significant simplification of the further calculations. At the same time, the presence of not only external momentum but also external frequency  does not allow to simply factorize subgraphs from a diagram and calculate them separately, as it was done in the six loop study of $\phi^4$ model \cite{KP16}.

In our renormalization scheme, the subtraction operation has the form
\begin{equation}
\overline\Gamma^R=(1-K)R'\overline\Gamma(k, \omega)=R'\overline\Gamma(k,\omega)-R'\overline\Gamma(k,\omega)|_{k=1, \omega=0}
\label{grsh}
\end{equation} 
that guarantees the fulfillment of the previously discussed condition $\overline \Gamma^R|_{k=1,\omega=0}=1$. The $K$-operation acts on the superficial divergent diagrams and does not affect unity in \eqref{gbar}. The  1-irreducible subgraphs $\sigma_j(k',\omega')=\langle vv'\rangle_{\text{1-irr}}(k',\omega')$ are the source of divergences, their divergent part is proportional to their external momentum squared $k'^2$. The subtraction operation acting on the normalized subgraph
\begin{equation}
\overline\sigma_j(k',\omega')=\frac{\sigma_j(k', \omega')}{k'^2},
\label{subp}
\end{equation} 
has the form analogous to \eqref{grsh} :
\begin{equation}
(1-K)\overline\sigma_j(k,\omega)=\overline\sigma_j(k,\omega,)-\overline\sigma_j(k=1,\omega=0).
\label{ksub}
\end{equation} 
Considering the independence of the quantity $R'\overline \Gamma(k,\omega )|_{k=1,\omega=0}$ on $k$, the relation \eqref{gnu3} can be rewritten as 
\begin{equation}
\gamma_\nu= {\cal L}(\overline\Gamma) = {\cal D}_k R'\overline\Gamma \equiv -(k\partial_k R'\overline\Gamma)|_{k=1,\omega=0}.
\label{gnu4}
\end{equation}  
In this relation, the perturbation series \eqref{gbar} for $\overline\Gamma(k,\omega=0)$, in terms of the introduced normalized parameters, is given by

\begin{equation}
\overline{\Gamma}(k,\omega=0) = 1 + \sum_{n\ge 1} h^n  \sum_{i} \chi_n^{(i)}(k,\omega=0).
\label{gbar2}
\end{equation}
 Substituting the expression \eqref{gbar2} into \eqref{gnu4} gives
\begin{equation}
\gamma_\nu=  \sum_{n\ge 1} h^n  \sum_{i}{\cal D}_k \left(R'\chi_n^{(i)}(k,\omega=0)\right).
\label{gnu5}
\end{equation} 

To apply the hyperlogarythm approach for the calculation of $\gamma_\nu$ in \eqref{gnu5}, one needs to obtain the expression for the diagrams, which determine the coefficients $\chi_n^{(i)}(k,\omega=0)$ in terms of the  Feynman parameter integrals.
As a first step, let us write the diagrams in the $k$-representation performing integration over $t$ in $(k,t)$-representation. It can be easily done using \eqref{l1}--\eqref{l3}. After that, only momenta integrals remain.

The important attribute of these integrals is that besides being dependent on the dimension of the space $d$ they also depend on the parameter $\ep$, which does not relate to $d$ in the considered model. The substantial simplifications appear when we consider limit $d \to \infty$. This can be achieved by making the optimal choice of the integration variables. Considering the $n$-loop diagram in the momentum representation, the number of lines $\langle vv\rangle$ coincides with the number of the integration momenta, being equal to $n$. Therefore, it is always possible to associate the simple integration momenta with the momenta on these lines. Then, in the spherical coordinate system, the dependence of the propagator \eqref{l1} on the dimension $d$ and the volume element factor $k^{d-1}$ cancel each other. Since in the limit $d \to \infty$ all the internal integration momenta are in fact orthogonal to each other, the integrands do not depend on the angles. The result of the angular integrations in spherical coordinate system is given by a factor $S_d$, which, along with the factor $1/(2\pi)^d$, is included in the definition \eqref{gbar} of the charge {
$h={g S_d }/{(2\pi)^d}$~\cite{AAGKK08}, factor $k^{d-1}$ from Jacobian along with factor $k^{2-d-2\varepsilon}$ from \eqref{l1} gives $k^{1-2\varepsilon}$}. The remaining integrals with no dependence on $d$ have the form 
{
\begin{equation}
\frac{g}{(2\pi)^d}\int d\vec{k} \;k^{2-d-2\varepsilon}...=h\int_0^\infty dk \; k^{1-2\ep}...
\label{kd}
\end{equation} 
}
The application of the hyperlogarithm method requires transforming obtained integrals to the parametric Feynman representation. 
They should be interpreted as ones in the dimension of the space equal to
\begin{equation}
D=2-2\ep,
\label{de}
\end{equation} 
with the integrands independent of angles. Such integrals can already be transformed to the Feynman representation using the standard, well-known technique.  It is important to emphasize that at $\ep=0$ the dimension \eqref{de} is even, which satisfies another condition for the applicability of the hyperlogarithm method.

The consideration of the $d \to \infty$ asymptotics leads to a significant reduction of the number of non-trivial diagrams. The main reason is that all diagrams with a scalar product of two different momenta are equal to zero and can be omitted. Such zero diagrams can be identified by analyzing their dynamic topologies; corresponding rules were formulated in \cite{AAGKK09_turb_3l}. Using them, it is possible to filter out all trivial zero diagrams, e.g., in 4 loops there are only 1693 significant diagrams of all 417872 possible.

Let us now discuss one of the calculated 3-loop diagrams in greater detail to illustrate the overall concept.
\begin{equation}
    \includegraphics[width=0.5\textwidth]{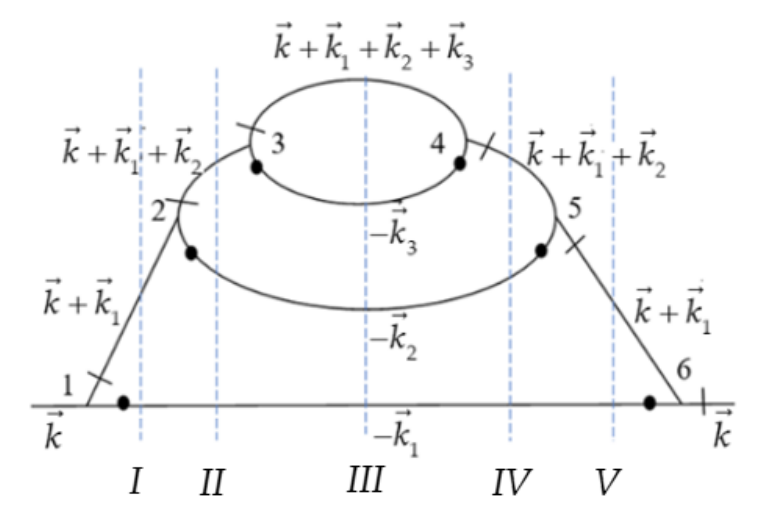} 
\label{dia3l}
\end{equation}
Here $\vec{k}$ is the external momentum. The integration momenta $\vec{k}_1, \vec{k}_2, \vec{k}_3$ correspond to the propagators $\langle vv\rangle$, in accordance with the general rule mentioned above.
{
Let us write the expression for this diagram for function $\overline{\Gamma}(k,\omega=0)$~\eqref{gch} in the $(k,t)$ representation. Using \eqref{l1}-\eqref{vertex}, we obtain \cite{AIKV_2018_4l}:
{\small
\begin{eqnarray}
\begin{aligned}
& J(\vec{k}, \omega=0)=-\frac{1}{\nu k^2} \cdot\left(\frac{g \mu^{2 \varepsilon} \nu^2}{2}\right)^3 \frac{1}{(2 \pi)^{3 d}} \int \frac{d \vec{k}_1}{k_1^{d-2+2 \varepsilon}} \int \frac{d \vec{k}_2}{k_2^{d-2+2 \varepsilon}} \int \frac{d \vec{k}_3}{k_3^{d-2+2 \varepsilon}}\left(-k^2\left(k^2+k_1^2\right)\left(k^2+k_1^2+k_2^2\right)\right) \cdot T \\
& T=\int_{-\infty}^{\infty} d t_1 \int_{-\infty}^{\infty} d t_2 \int_{-\infty}^{\infty} d t_3 \int_{-\infty}^{\infty} d t_4 \int_{-\infty}^{\infty} d t_5 \,\, \theta\left(t_2-t_1\right) \theta\left(t_3-t_2\right) \theta\left(t_1-t_3\right) \theta\left(t_5-t_4\right) \theta\left(t_6-t_1\right) \cdot \\
& \cdot \exp \left\{-\nu\left[\begin{array}{l}
\left(\vec{k}+\vec{k}_1\right)^2\left(t_2-t_1\right)+\left(\vec{k}+\vec{k}_1+\vec{k}_2\right)^2\left(t_3-t_2\right)+\left(\vec{k}+\vec{k}_1+\vec{k}_2+\vec{k}_3\right)^2\left(t_1-t_3\right)+ \\
\left(\vec{k}+\vec{k}_1+\vec{k}_2\right)^2\left(t_5-t_4\right)+\left(\vec{k}+\vec{k}_1\right)^2\left(t_6-t_5\right)+k_3^2\left|t_4-t_3\right|+k_2^2\left|t_5-t_2\right|+k_1^2\left|t_6-t_1\right|
\end{array}\right]\right\}
\label{J}
\end{aligned}
\end{eqnarray}
}
The factors in the numerator come from the contraction of the momenta corresponding to the bold dots. It is taken into account that the squared sum of the momenta reduces to the sum of the squared momenta due to neglecting the scalar product of any non-collinear vectors. The time $t_6$ is considered fixed, which corresponds to the condition $\omega=0$. Since $J$ obviously does not depend on $\nu$, we will henceforth set $\nu=1$. Thus, $E_k=\nu k^2$ becomes $E_k=k^2$. For calculating $T$, one could transition to relative times: $\tau_1 = t_2-t_1$, $\tau_2 = t_3-t_2$, $\tau_3 = t_4-t_3$, $\tau_4 = t_5-t_4$, $\tau_5 = t_6-t_5$. Then, the exponents $t_5-t_2$ and $t_6-t_1$ will be expressed in terms of $\tau_i$. In these variables, the time integrals factorize and are easy to evaluate in the limits $(0,\infty)$. As a result, we obtain
\begin{eqnarray}
    T\left(\vec{k},\vec{k_1},\vec{k_2},\vec{k_3}\right) = \frac{1}{Q\left(k,k_1,k_2,k_3\right)} \,\, ,
\end{eqnarray}
where $Q$ is defined as follows
\begin{eqnarray}
Q(k,k_1,k_2,k_3)=(E_{k_1}+E_{k+k_1})(E_{k_1}+E_{k_2}+E_{k+k_1+k_2})(E_{k_1}+E_{k_2}+E_{k_3}+E_{k+k_1+k_2+k_3})\cdot \nonumber\\
\cdot(E_{k_1}+E_{k_2}+E_{k+k_1+k_2})(E_{k_1}+E_{k+k_1})=\Big[k_1^2+(k^2+k_1^2)\Big]\Big[k_1^2+k_2^2+(k^2+k_1^2+k_2^2)\Big]\cdot \nonumber\\\cdot\Big[k_1^2+k_2^2+k_3^2+(k^2+k_1^2+k_2^2+k_3^2)\Big]
\Big[k_1^2+k_2^2+(k^2+k_1^2+k_2^2)\Big]\Big[k_1^2+(k^2+k_1^2)\Big].
\end{eqnarray}
This result can be interpreted as the multiplication of 5 so-called "time-cuts" \eqref{dia3l}
\begin{eqnarray}
Q(k,k_1,k_2,k_3)= {\cal E}_I \; {\cal E}_{II}\; {\cal E}_{III} \;{\cal E}_{IV}\; {\cal E}_V \,,
\label{Q}
\end{eqnarray}
where ${\cal E}_i= \sum_n E_n$ and $n$ represents for propagators intersecting the dashed line.
}

{
The general approach to performing integration over times involves two steps. First, one needs to arrange vertices of a diagram in all possible sequences with respect to $\theta$-functions present in $\left<vv'\right>$ propagators. These diagrams with ordered sequence of vertices are called "time versions". Then, in each of them one needs to place time-cuts between all adjacent pairs of vertices, which would represent integration over relative times. The result of such integration is equal to {\it$1/$(sum of squared momenta, intersected with time-cut)}. This algorithm, being general for dynamic models, is well-suited for automation, making the task of integrating over times straightforward.}
{

Taking out factor $h^3\mu^{-6\varepsilon}$ from $J(\vec{k}, \omega=0)$~\eqref{J} 
\begin{equation}
    J(\vec{k}, \omega=0)=   h^3\mu^{-6\varepsilon} \chi(k,\omega=0)\,,
\end{equation}
which along with \eqref{gbar} from \eqref{J}-\eqref{Q} and \eqref{kd} gives
\begin{eqnarray}
\chi(k)=\chi(k,\omega=0)=\frac{1}{8}\int_0^\infty dk_1 k_1^{1-2\ep}\int_0^\infty dk_2 k_2^{1-2\ep}\int_0^\infty dk_3 k_3^{1-2\ep}\frac{(k^2+k_1^2)(k^2+k_1^2+k_2^2)}{Q(k,k_1,k_2,k_3)}\;.
\end{eqnarray} 
}


Now, let us discuss the implementation of the $R'$-operation written in the form \eqref{Rzav}. To perform the $(1-K)_j$ operation on the relevant subgraph with the index $j$, one can introduce the parameter $a_j$ and put it into the integrand in such a way that the integrand remains the same if $a_j=1$ and the value $a_j=0$ corresponds to the values $\omega_j=0, k_j=1$ of the external frequency and momentum. The $(1-K)_j$ operation ~\eqref{ksub} is represented as the difference between the diagram expression at $a_j=1$ and $a_j=0$. Then, the result of the $R'$-operation is provided by such substitutes for all divergent subgraphs.

The formulated conditions for the parameter $a_j$ can be implemented in the following way. To satisfy the condition $\omega_j=0$ in the cuts crossing the subgraph one needs to introduce the factor $a_j$ before the energies not containing the integration variables of the subgraph. For condition $k_j=1$, one should replace $k_j$ with $k_j^{a_j}$ in all other energies. The same replacement needs to be done in the numerator factors that contain the integration variables of the subgraph.

Let us apply the formulated rules to the diagram \eqref{dia3l}. It has a subgraph with vertices $\{3,4\}$, which is nested to another one with vertices $\{2,3,4,5\}$. The first subgraph is associated with the parameter $a_1$ and the second one is associated with $a_2$. The first subgraph ($\{3,4\}$) has only one cut in which the energies $E_{k_1}$ and $E_{k_2}$ do not depend on the subgraph integration variable $k_3$, therefore one needs to multiply them by the factor $a_1$. The external momentum is equal to $\mathbf{k}+\mathbf{k}_1+\mathbf{k}_2$, thus in the energy $E_{k+k_1+k_2+k_3}$ the combination $k^2+k_1^2+k_2^2$ is replaced with $(k^2+k_1^2+k_2^2)^{a_1}$. There is no need to manipulate the factor $k^2+k_1^2+k_2^2$ in the numerator because the subtraction is applied to the normalized subgraph \eqref{subp}. The second subgraph ($\{2,3,4,5\}$) has the external momentum $\mathbf{k}+\mathbf{k}_1$ and contains three cuts. Each of them contains the energy $E_{k_1}$, which needs to be multiplied by the factor $a_2$ due to its independence of the integration variable of the subgraph. In the rest of the energies, the factor $k^2+k_1^2$ is replaced with $(k^2+k_1^2)^{a_2}$. The same replacement is needed for the factor $k^2+k_1^2+k_2^2$ in the numerator. The final result is given by

\begin{equation}
\chi_{a_1,a_2}(k)=\frac{1}{8}\int_0^\infty dk_1 k_1^{1-2\ep}\int_0^\infty dk_2 k_2^{1-2\ep}\int_0^\infty dk_3 k_3^{1-2\ep}\frac{(k^2+k_1^2)\big[(k^2+k_1^2)^{a_2}+k_2^2\big]}{Q_{a_1,a_2}(k,k_1,k_2,k_3)}
\label{waa}
\end{equation} 
\begin{eqnarray}
Q_{a_1,a_2}=\Big\{k_1^2+(k^2+k_1^2)\Big\}\Big\{a_2k_1^2+k_2^2+\Big[(k^2+k_1^2)^{a_2}+k_2^2\Big]\Big\}\cdot\nonumber\\\cdot\Big\{a_1a_2k_1^2+a_1k_2^2+k_3^2+\Big[\Big((k^2+k_1^2)^{a_2}+k_2^2\Big)^{a_1}+k_3^2\Big]\Big\}\Big\{
\Big[(k^2+k_1^2)^{a_2}+k_2^2\Big]\Big\}\Big\{k_1^2+(k^2+k_1^2)\Big\}.
\label{qaa}
\end{eqnarray}
Substitution of the parameter values yields the result of the $R'$-operation in the form
\begin{equation}
\label{R'_exmpl}
R'\chi(k)=\chi_{1,1}(k)-\chi_{1,0}(k)-\chi_{0,1}(k)+\chi_{0,0}(k).
\end{equation} 
Thus, the ${\cal L}$ operation from \eqref{gnu4} is given by
\begin{eqnarray}
\label{L_exmpl}
{\cal L}(\chi(k))&=&{\cal D}_k (R'\chi(k))= (-k\partial_kR'\chi(k))|_{k=1}=\\
&=&{\cal D}_k(\chi_{1,1}(k))-{\cal D}_k(\chi_{1,0}(k))-{\cal D}_k(\chi_{0,1}(k))+{\cal D}_k(\chi_{0,0}(k)).
\nonumber
\end{eqnarray}

Let us stress that in the case of nested subgraphs, the factors $a_i$ must be arranged starting with the smallest corresponding subgraph as in the considered diagram. It is important to emphasize once again that there are no overlapping subgraphs in this model. 

In the following, we discuss the transition to the  Feynman representation. In the general case of an $n$-loop diagram after performing all integrations over times the integrand contains in the denominator $2n-1$ factors $A_i$ quadratic in momenta  and in the numerator $n-1$ factors $B_i$ quadratic in momenta as well. In this context, the Feynman formula is written as follows 
\begin{equation}
\frac{B_1...B_{n-1}}{A_1..A_{2n-1}}=(n-1)!\left[(-\partial_{u'_1})...(-\partial_{u'_{n-1}})\int_0^1 du_1...\int_0^1 du_{2n-1}\frac{\delta\Big(\sum\limits_{i=1}^{2n-1}u_i-1\Big)}{\Big(\sum\limits_{i=1}^{2n-1}A_iu_i+\sum\limits_{i=1}^{n-1}B_iu'_i\Big)^n}\right]\Bigg|_{u'_1=0,\ldots ,u'_{n-1}=0}
\label{Feynman}
\end{equation}
\noindent Here,
\begin{equation}
H\equiv\Big(\sum\limits_{i=1}^{2n-1}A_iu_i+\sum\limits_{i=1}^{n-1}B_iu'_i\Big) =\sum\limits_{i,s=1}^n v_{is}k_ik_s+c
\label{H}
\end{equation} 
is the quadratic form of the integration variables. Considering the integrals over modules of $k_i$ as the integrals in $D$-dimensional space and taking into account that the integrands are independent of angles
\begin{equation}
\int_0^\infty k_1^{1-2\ep} dk_1...=\frac{1}{S_D}\int d \mathbf{k_1} ..., \quad S_D=\frac{2\pi^{D/2}}{\Gamma(D/2)}, \quad D=2-2\ep
\label{i1}
\end{equation} 
one can use the formula for the integration of quadratic form raised to some power:
\begin{equation}
\frac{1}{S_D^n}\int d\mathbf{k_1} ...\int d \mathbf{k_n} \frac{1}{H^n}=\frac{\Gamma^n(D/2)\Gamma(n-nD/2)}{2^n(n-1)!c^{n-nD/2}{\cal U}^{D/2}}=\frac{\Gamma^n(1-\ep)\Gamma(n\ep)}{2^n(n-1)!c^{n\ep}{\cal U}^{1-\ep}},
\label{i2}
\end{equation} 
where we denote ${\cal U} = \det v$. With \eqref{Feynman}-\eqref{i2}, the general formula of the Feynman representation is given by
\begin{eqnarray}
\label{final_repr}
\int_0^\infty k_1^{1-2\ep}dk_1...\int_0^\infty k_2^{1-2\ep} dk_n \frac{B_1...B_{n-1}}{A_1...A_{2n-1}}=\frac{1}{S_D^n}\int d \mathbf{k_1}...\int d\mathbf{k_n}  \frac{B_1...B_{n-1}}{A_1...A_{2n-1}}=\nonumber\\
=\frac{\Gamma^n(1-\ep)\Gamma(n\ep)}{2^n}\left[(-\partial_{u'_1})...(-\partial_{u'_{n-1}})\int_0^1 du_1...\int_0^1 du_{2n-1}\frac{\delta\Big(\sum\limits_{i=1}^{2n-1}u_i-1\Big)}{c^{n\ep}{\cal U}^{1-\ep}}\right]\Bigg|_{u'_1=0...u'_{n-1}=0}
\end{eqnarray} 
In this model, the matrix $v$ is always diagonal, which is a direct consequence of the vanishing of the scalar products. 
\begin{equation}
v_{is}=v_i\delta_{is}, \quad {\cal U}=\prod_{i=1}^n v_i \;.
\end{equation} 
This fact leads to significant simplifications in the calculations. Denoting the quadratic form \eqref{H} of the diagram \eqref{dia3l} for various values of the parameters $a_i$ as $H_{a_1a_2}$, one obtains
\begin{eqnarray}
H_{1,1}=(k^2+2k_1^2)u_1+(k^2+2k_1^2+2k_2^2)u_2+(k^2+2k_1^2+2k_2^2+2k_3^2)u_3+(k^2+2k_1^2+2k_2^2)u_4+\nonumber\\+(k^2+2k_1^2)u_5+(k^2+k_1^2)u'_1+(k^2+k_1^2+k_2^2)u'_2=k^2(u_1+u_2+u_3+u_4+u_5+u'_1+u'_2)+\\+k_1^2(2u_1+2u_2+2u_3+2u_4+2u_5+u'_1+u'_2)+k_2^2(2u_2+2u_3+2u_4+u'_2)+k_3^22u_3,\nonumber
\end{eqnarray} 
that yields
\begin{equation}
\label{c11}
c_{1,1}=k^2(u_1+u_2+u_3+u_4+u_5+u'_1+u'_2),
\end{equation} 
\begin{equation}
\label{u11}
{\cal U}_{1,1}=(2u_1+2u_2+2u_3+2u_4+2u_5+u'_1+u'_2)(2u_2+2u_3+2u_4+u'_2)2u_3,
\end{equation} 
and, analogically,
\begin{equation}
c_{1,0}=k^2(u_1+u_5+u'_1)+u_2+u_3+u_4+u'_2,
\end{equation} 
\begin{equation}
{\cal U}_{1,0}=(2u_1+2u_5+u'_1)(2u_2+2u_3+2u_4+u'_2)2u_3,
\end{equation} 
\begin{equation}
c_{0,1}=k^2(u_1+u_2+u_4+u_5+u'_1+u'_2)+u_3,
\end{equation} 
\begin{equation}
{\cal U}_{0,1}=(2u_1+2u_2+2u_4+2u_5+u'_1+u'_2)(2u_2+2u_4+u'_2)2u_3,
\end{equation} 
\begin{equation}
c_{0,0}=k^2(u_1+u_5+u'_1)+u_2+u_3+u_4,
\end{equation} 
\begin{equation}
{\cal U}_{0,0}=(2u_1+2u_5+u'_1)(2u_2+2u_4+u'_2)2u_3.
\end{equation}

Finally, the resulting Feynman representation for the coefficients $\chi^{(i)}_n(k,\omega=0)$ in the relation \eqref{gnu5} allows to apply the hyperlogarithm technique for calculating the anomalous dimension $\gamma_\nu$.

\section{Calculations with the hyperlogarithm method} \label{HL_sec}
For evaluation of the diagrams, we use a hyperlogarithm integration method, which is a powerful tool for multiloop calculations. It was proposed by Francis Brown in the series of works \cite{Brown08, Brown09, Brown09_arx} and then further developed by Erik Panzer~\cite{Panzer14_main, Panzer_thesis, Panzer14_diverg}, resulting in a Maple implementation \textit{HyperInt}, which we apply to perform analytic  calculations of the diagrams.

The hyperlogarithm method is applicable for calculation of diagrams without divergences. Since it is not directly suitable for divergent diagrams, one needs to construct some combination ${\cal L}(\chi)$ in which the divergences cancel out on the level of the integrand. To calculate the initial integral, one needs to calculate the sum ${\cal L}(\chi)$ and the subtraction terms. The corresponding subtractions tend to be easier to evaluate than the original diagram due to the simplifications of the integrands, e.g., factorization. In our case, the operation ${\cal L}$ in \eqref{gnu4}  is simpler  because it is constructed on the basis of the standard $R'$ operation, and the value ${\cal L}(\chi)$ directly contributes to the required RG function $\gamma_\nu$~\eqref{gnu4}.  

 Let's have a detailed look at the hyperlogarithm method. The basic object of the discussed method is a hyperlogarithm, which is defined in terms of iterated integrals as
\begin{equation}
\label{hlog1}
L_{\omega_{\sigma_1}\omega_{\sigma_2}...\omega_{\sigma_n}}(z) := \int_0^{z}\frac{dz_1}{z_1-\sigma_1}\int_0^{z_1}\frac{dz_2}{z_2-\sigma_2}...\int_0^{z_{n-1}}\frac{dz_n}{z_n-\sigma_n} \,\,,
\end{equation} 
 where the letter $\omega_{\sigma_i}$ is the symbol for the differential form ${dz_i}/({z_i-\sigma_i})$. The combination of letters in the definition (\ref{hlog1}) is called the word $w=\omega_{\sigma_1}\omega_{\sigma_2}...\omega_{\sigma_n}$. The one consisting of the same letter repeated $\underbrace{\omega_{\sigma_i}\,...\,\omega_{\sigma_i}}_n$ will be denoted as $\omega_{\sigma_i}^n$.

For the application of the hyperlogarithm technique, it is necessary to use the well-known trick in the parametric representation (\ref{final_repr}) that is often referred to as the Cheng-Wu theorem~\cite{chengwu}. In particular, it allows to replace the $\delta$-function $\delta\left(\sum_{i=1}^{n}u_i-1\right)$ with $\delta\left(u_j-1\right)$ for an arbitrary parameter $u_j$ by having the limits of integration over the remaining parameters extended to infinity. Due to the finiteness of the calculated expression ${\cal L}(\chi)$ ((\ref{L_exmpl}) in the example considered above), its $\varepsilon$-expansion can be obtained by expanding the integrand into a Taylor series at $\varepsilon=0$. The individual terms in ${\cal L}(\chi)$ (see, e.g, (\ref{L_exmpl})) have logarithmic divergences, which cancel out completely. Being the core element of the hyperlogarithm method, the regularization procedure allows for the evaluation of each term separately. It is a drastic simplification compared to the numeric calculation, which requires combining all the terms into one finite expression.

After the subsequent integration over Feynman parameters, the calculated integrals can be represented in terms of the regularized limits of hyperlogarithms $\underset{z= \infty}{\textrm{Reg}}L_w(z)$ of the words $w$ with letters $\omega_0 (\sigma=0)$ and $\omega_{-1} (\sigma=-1)$. The basic examples of the regularization of such hyperlogarithms are
\begin{equation}
\label{reg_exmpl1}
\underset{z= \infty}{\textrm{Reg}}L_{\omega^n_{-1}}(z) = \frac{1}{n!}\underset{z= \infty}{\textrm{Reg}}\ln^n(1+z) = \frac{1}{n!}\underset{z= \infty}{\textrm{Reg}}\left[ \ln^n(z) + \mathcal{O}\left( \frac{1}{z} \right) \right] = 0\,\,,
\end{equation} 
\begin{equation}
\label{reg_exmpl2}
\underset{z= \infty}{\textrm{Reg}}L_{\omega_0\omega^n_{-1}}(z) = L_{\omega_0\omega^n_{-1}}(z) - L_{\omega^{n+1}_{-1}}(z) = L_{(\omega_0-\omega_{-1})\omega^n_{-1}}(\infty)\,\,.
\end{equation} 
The first letter $\omega_0-\omega_{-1}$ in (\ref{reg_exmpl2}) corresponds to the expression ${dz_1}(1/{z_1} - 1/({z_1+1})) = {dz_1}/({z_1(z_1+1)})$ in the definition (\ref{hlog1}), which ensures the convergence of the integral at $z \xrightarrow{} \infty$. 

In the general case, the subtraction of this type leads to the convergent integral, but, unlike in (\ref{reg_exmpl1}), $\underset{z= \infty}{\textrm{Reg}} L_{\omega_{-1} \omega_{\sigma_2} ... \omega_{\sigma_n}}(z) \neq 0$, so the finite part must be eliminated from the subtractive term.
 The sequential procedure for constructing the regularized limit of an arbitrary word $\underset{z= \infty}{\textrm{Reg}}L_{\omega_{\sigma_1}\omega_{\sigma_2}...\omega_{\sigma_n}}(z)$ is discussed in \cite{Panzer_thesis} in great detail. Let us only provide the following example 
\begin{equation}
\label{reg_exmpl3}
\underset{z= \infty}{\textrm{Reg}}L_{\omega_0\omega_0\omega_{-1}\omega_{-1}}(z) = L_{(\omega_0-\omega_{-1})\omega_0\omega_{-1}\omega_{-1}}(\infty) - 3L_{(\omega_0-\omega_{-1})\omega_{-1}\omega_{-1}\omega_{-1}}(\infty)\,\,.
\end{equation}

 The regularized hyperlogarithms with the letters $\omega_{0}$ and $\omega_{-1}$ can be expressed in terms of multiple zeta functions. The Riemann zeta function of one argument is defined as
\begin{equation}
\label{zeta_basic_def}
\zeta(n) = \sum_{k=1}\frac{1}{k^n}\,\,.
\end{equation}
 Its integral representation corresponds to the value of the polylogarythm at $z=1$
\begin{equation}
\zeta(n) = polylog(n,z=1)=-\int^{z=1}_0\frac{dt_1}{t_1} \int_0^{t_1}\frac{dt_2}{t_2}...\int^{t_{n-2}}_0\frac{dt_{n-1}}{t_{n-1}}\ln(1-t_{n-1}),
\end{equation}
 or, in terms of hyperlogarithms,
\begin{equation}
\label{zeta_def1}
\zeta(n) = \int^{1}_0\frac{dt_1}{t_1} \int_0^{t_1}\frac{dt_2}{t_2}...\int^{t_{n-2}}_0\frac{dt_{n-1}}{t_{n-1}}\int_0^{t_{n-1}}\frac{dt_{n}}{t_n-1} = L_{\omega_0^{n-1}\omega_1}(1)\,\,.
\end{equation}
 Multiple zeta function is defined as
\begin{equation}
\zeta(n_1,n_2,...,n_m) = \sum_{0<k_1<k_2<...<k_m}\frac{1}{k_1^{n_1} k_2^{n_2} ... k_m^{n_m}}\,\,,
\end{equation}
 its integral representation can be obtained from (\ref{zeta_def1}) by inserting certain amount of letters $\omega_1$, namely,
\begin{equation}
\label{zeta_via_reg}
\zeta(n_1,n_2,...,n_m) = (-1)^mL_{\omega_0^{n_m-1}\omega_1...\omega_0^{n_2-1}\omega_1\omega_0^{n_1-1}\omega_1}(1)\,\,.
\end{equation}

 The hyperlogarithms in (\ref{zeta_via_reg}) can be constructed from the regularized limits $\underset{z= \infty}{\textrm{Reg}}L_{\omega_{\sigma_1}\omega_{\sigma_2}...\omega_{\sigma_n}}(z)$ with letters $\omega_0$ and $\omega_{-1}$ with the following change of variables  
\begin{equation}
z_i = \frac{t_i}{1-t_i}\,,\,\,\,\,\,\,\frac{dz_i}{z_i} = \frac{dt_i}{t_i}+\frac{dt_i}{1-t_i}\,,\,\,\,\,\,\,\frac{dz_i}{z_i+1} = \frac{dt_i}{1-t_i}\,\,,
\end{equation}
 which corresponds to the change of letters
\begin{equation}
\label{var_change1}
\omega_0 \,\,\,\xrightarrow{\,\,\,\,\,\,}\,\,\omega_0 - \omega_1\,\,,\,\,\,\,\,\,\,\,\,\,\,\,\,\,\,\omega_{-1} \xrightarrow{\,\,\,\,\,\,} \,\,-\omega_1\,\,.
\end{equation}
 Let us note that the first letter in the regularized hyperlogarithm will always be $\omega_0 - \omega_{-1}$ (as in the example (\ref{reg_exmpl3})), ensuring the existence of the limit at $z \xrightarrow{} \infty$. Consequently, after applying (\ref{var_change1}) the first letter in (\ref{zeta_via_reg}) will be $\omega_0$, which leads to the inequality $n_m\geq2$, preventing the occurrence of the divergent $\zeta(n_1,n_2,...,1)$.
 
 It is known that multi zeta values with weights (sum of the arguments) up to and including $7$ can be expressed as the sum of products of Riemann zeta values $\zeta(n)$. In particular, for the regularization examples (\ref{reg_exmpl2}),(\ref{reg_exmpl3}) the formulae (\ref{zeta_via_reg}), (\ref{var_change1}) yield
\begin{equation}
\label{reg_0_-1}
\underset{z= \infty}{\textrm{Reg}}L_{\omega_0\omega^n_{-1}}(z) = L_{(\omega_0-\omega_{-1})\omega^n_{-1}}(\infty) = (-1)^n L_{\omega_0\omega^n_{1}}(1) = (-1)^n \zeta(\underbrace{1,1,...,1}_n,2) = \zeta(n+1)\,\,\,\,\,\,\,\,\,\,\,\,\,\,\,
\end{equation}
\begin{align}
&\underset{z= \infty}{\textrm{Reg}}L_{\omega_0\omega_0\omega_{-1}\omega_{-1}}(z) = L_{(\omega_0-\omega_{-1})\omega_0\omega^2_{-1}}(\infty) - 3L_{(\omega_0-\omega_{-1})\omega^3_{-1}}(\infty)
=L_{\omega_0(\omega_0-\omega_{1})\omega^2_1}(1) + 3L_{\omega_0\omega^3_1}(1) =\nonumber\\
&= L_{\omega^2_0\omega^2_1}(1) + 2L_{\omega_0\omega^3_1}(1) =  \zeta(1,3) - 2\zeta(1,1,2) = \frac{1}{10}\zeta^2(2)-\frac{4}{5}\zeta^2(2)=-\frac{7}{10}\zeta^2(2)
\end{align} 
Therefore, all considered multiple integrals can be reduced to the regularized limits of the hyperlogarithms of words with the letters $\omega_0$ and $\omega_{-1}$. For illustration, we provide the calculation of the first term ${\cal D}_k(\chi_{1,1}(k))$ in the sum \eqref{L_exmpl} corresponding to the diagram (\ref{dia3l}). 
The general integration algorithm, which is quite complex, can be found in \cite{Panzer_thesis}.

{
Consider the term $D_k (\chi_{1,1})$ at $\varepsilon=0$, using the representation \eqref{final_repr} with $n=3$. From \eqref{final_repr},\eqref{c11},\eqref{u11} it follows that $\chi_{1,1}(k)=k^{-6\varepsilon}\chi_{1,1}(1)$. Therefore, $D_k\chi_{1,1}(k)=6\varepsilon\chi_{1,1}(k)$. The factor $6\varepsilon$ cancels the pole in the factor ${\Gamma^3(1-\varepsilon)\Gamma(\varepsilon)}/{2^3}$ from \eqref{final_repr}, resulting in $6\ep {\Gamma^3(1-\varepsilon)\Gamma(\varepsilon)}/{2^3}={1}/{4}+\mathcal{O}(\varepsilon)$. Substituting \eqref{c11},\eqref{u11} into \eqref{final_repr} and taking the derivatives {\small$\partial_{u'_1} \partial_{u'_2}...\Big|_{u'_1=u'_2=...=0}$}, we finally obtain
}

\begin{equation}
{\cal D}_k(\chi_{1,1}(k)) = \frac{1}{1024}\int_{0}^{1}du_1 \ldots  du_5 \frac{(u_1+3u_2+3u_3+3u_4+u_5)\,\delta(u_1+u_2+u_3+u_4+u_5-1)}{u_3\,(u_2+u_3+u_4)^2\,(u_1+u_2+u_3+u_4+u_5)^3}
\end{equation}
To deal with the $\delta$-function as discussed in the beginning of the section, let us set $u_3=1$  and expand the integration limits over the other parameters to $\infty$
\begin{equation}
\label{first_term}
{\cal D}_k(\chi_{1,1}(k))= \frac{1}{1024}\int_{0}^{\infty}du_2 du_4 du_5 du_1 \frac{(u_1+3u_2+3u_4+u_5+3)}{(u_2+u_4+1)^2\,(u_1+u_2+u_4+u_5+1)^3}
\end{equation}
For convenience, the integration order will be the following: $u_2,u_4,u_5,u_1$. Diagrams of our model are linearly reducible for all orders of integration, we only note that the integration order must be chosen consistently for all terms of the diagram since they are divergent objects and have to be considered together.

{
To compute an integral with the hyperlogarithm method, the requirement of linear reducibility must be satisfied. This guarantees that at each step of integration, it is possible to construct a primitive in terms of hyperlogarithms, and the final result will be expressed as a rational linear combination of multiple zeta values. In general, an integrand at some integration step has the form of a product of rational function and hyperlogarithm. An integral is considered linearly reducible if, at each step, the irreducible polynomials in the rational functions remain linear in the next integration variables. For some integrals, linear reducibility is not satisfied for an arbitrary integration order, so finding a proper one is mandatory. Several reduction algorithms exist that can be used to search for integration orders which ensure linear reducibility, without actually doing the integration process \cite{Brown09, Panzer_thesis}. The presence of linear reducibility depends on diagrams' topology, but this topic has been studied for only static models, and even in those its nature is not yet completely understood. Fortunately, in the considered dynamic model all diagrams turn out to be linearly reducible up to 4 loops due to significant simplifications of the $d\to \infty$ limit. In this regard, we only note that the integration order must be chosen consistently for all terms of the diagram since they are divergent objects and must be treated together. For convenience, the integration order in the considered example (\ref{first_term}) will be the following: $u_2,u_4,u_5,u_1$.
}

Every integration except the last integral over $u_1$ can be taken by standard methods of partial fraction decomposition and integration by parts. The result of the integration over $u_2$ is denoted as $\chi_{1,1}^{(2)}$, where
\begin{equation}
{\cal D}_k(\chi_{1,1}(k)) = \frac{1}{1024} \int_0^{\infty} du_4 du_5 du_1 \, \chi_{1,1}^{(2)}\,\,,
\end{equation}

\begin{align}
&\chi_{1,1}^{(2)} = \int_0^{\infty} du_2 \left[ -\frac{2}{(u_1+u_5)\,(u_1+u_2+u_4+u_5+1)^3} - \frac{1}{(u_1+u_5)^2\,(u_1+u_2+u_4+u_5+1)^2}+ \right. \nonumber \\
& \left. + \frac{1}{(u_1+u_5)^2\,(u_2+u_4+1)^2} \right] =\frac{1}{(u_1+u_4+u_5+1)^2\,(u_4+1)} \;.\nonumber
\end{align}
Integrating over $u_4$ and $u_5$, we get 
\begin{align}
&\chi_{1,1}^{(2,4)} = \int_0^{\infty} du_4 \, \chi_{1,1}^{(2)} = \int_0^{\infty} du_4 \left[ \frac{1}{(u_1+u_5)^2\,(u_4+1)} - \frac{1}{(u_1+u_5)^2\,(u_1+u_4+u_5+1)} \right. - \,\,\,\,\,\,\,\,\,\,\,\,\,\,\nonumber \\
& - \left. \frac{1}{(u_1+u_5)\,(u_1+u_4+u_5+1)^2} \right] = \frac{\ln(u_1+u_5+1)}{(u_1+u_5)^2} - \frac{1}{(u_1+u_5)\,(u_1+u_5+1)}\;,
\end{align}

\begin{align}
\chi_{1,1}^{(2,4,5)} = \int_0^{\infty} du_5 \,\chi_{1,1}^{(2,4)} = \left[ \frac{\ln(u_1+1)}{u_1} + \ln(u_1+1) - \ln(u_1) \right] + \left[-\ln(u_1+1) + \ln(u_1) \right] = \frac{\ln(u_1+1)}{u_1}\;.
\end{align}

The resulting expression is logarithmically divergent when integrated over the remaining parameter $u_1$. Now we represent it as a hyperlogarithm and perform the regularization that eliminates the divergent part. From the definition (\ref{hlog1}) follows that
\begin{equation}
\label{main_int_rule1}
\int_a^{b} dz \, \frac{L_w(z)}{z-\sigma} = L_{\omega_\sigma w}(b) - L_{\omega_\sigma w}(a)\,\,.
\end{equation} 
Rewriting $\ln(u_1+1)$ as the hyperlogarithm $L_{\omega_{-1}}(u_1)$ and using (\ref{main_int_rule1}), we obtain
\begin{align}
\chi_{1,1}^{(2,4,5,1)} = \int_0^{\infty} du_1 \,\chi_{1,1}^{(2,4,5)} = \int_0^{\infty} du_1 \frac{L_{\omega_{-1}}(u_1)}{u_1} = \underset{u_1 = \infty}{\textrm{Reg}}\,L_{\omega_0\omega_{-1}}(u_1) - \underset{u_1 = 0}{\textrm{Reg}}\,L_{\omega_0\omega_{-1}}(u_1)\;.
\end{align}
 The integral is reduced to the regularized limits of the hyperlogarithm of word with the letters $\omega_0$ and $\omega_{-1}$ as expected. The regularization at $u_1=0$ corresponds to the zero upper limit in (\ref{hlog1}), so it is equal to zero. The regularization at $\infty$ can be obtained from (\ref{reg_0_-1}), thus
\begin{align}
&\underset{u_1 = \infty}{\textrm{Reg}}\,L_{\omega_0\omega_{-1}}(u_1) = \zeta(2)\,\,, \nonumber \\
&\underset{u_1 = 0}{\textrm{Reg}}\,L_{\omega_0\omega_{-1}}(u_1) = 0  \,\,.
\end{align}
Therefore, the complete answer for the considered term \eqref{first_term} is 
\begin{align}
{\cal D}_k(\chi_{1,1}) = \frac{\zeta(2)}{1024}\,\,.
\end{align} 
 Let us provide the results for the remaining terms from \eqref{L_exmpl}:
\begin{equation}
{\cal D}_k(\chi_{1,0}) = -\frac{3}{1024} + \frac{\zeta(2)}{1024}\,\,,\qquad{\cal D}_k(\chi_{0,1}) = -\frac{1}{2048}\,\,,\qquad{\cal D}_k(\chi_{0,0}) = -\frac{1}{512}\,\,.
\end{equation} 
 The sum of all four terms would give the final result for the diagram (\ref{dia3l})
\begin{equation}
{\cal L}(\chi) = {\cal D}_k(\chi_{1,1}(k))-{\cal D}_k(\chi_{1,0}(k))-{\cal D}_k(\chi_{0,1}(k))+{\cal D}_k(\chi_{0,0}(k)) = \frac{3}{2048}\,\,,
\end{equation} 
 which, along with the other three-loop diagrammatic contributions, will be included in the sum (\ref{gnu5}).

 The considered example is relatively simple: there was no need to introduce any additional techniques up to the last integration over $u_1$, where the integral was represented in terms of the regularized hyperlogarithms of the word with letters $\omega_0$ and $\omega_{-1}$. In many other diagrams, the integration process may be much more sophisticated because hyperlogarithms may arise in the earlier stages which leads to regularization of words with letters, consisting of combinations of other integration parameters. To automatically perform these cumbersome calculations for all diagrams up to 4 loops included, we used the already mentioned implementation of this algorithm called \textit{HyperInt} \cite{Panzer14_main}, which allows for obtaining analytic results for Feynman integrals of different types. 
It should be noted that calculations were performed with \textit{HyperInt} option $\_hyper\_check\_divergences:=false$, otherwise you will be unable to calculate terms (in our example ${\cal D}_k(\chi_{1,1})$, ${\cal D}_k(\chi_{1,0})$, ${\cal D}_k(\chi_{0,1})$,  and ${\cal D}_k(\chi_{0,0})$) separately.

\section{Results and discussion}

To evaluate the RG function (\ref{gnu5}), we developed a program~\cite{repo_turb} that automates all computation steps, namely, generating diagrams, selecting significant ones (that do not contain scalar products), integrating over times, transforming to parametric representation, and constructing $R'$-operation as described in the sections above. Integration with the hyperlogarithm method was performed using the program \textit{HyperInt} \cite{Panzer14_main}, which has proven itself in multiloop calculations. Computations in each order of perturbation theory were carried out with consistent accuracy, taking into account the fact that at the fixed point of the renormalization group, the charge $u_* \sim \varepsilon$.

A result is written more compactly in terms of charge {
${\cal H}=2^{\ep-2}h$}: 
{
\begin{eqnarray}
\label{an_dim_ans}
\gamma_\nu={\cal H}\left (1+\frac{\pi^2\ep^2}{6} + 0 \cdot \ep^3\right )+{\cal H}^2\left (\frac{1}{2}-\frac{(\pi^2+9)\ep}{6}+\frac{(\pi^2+8\zeta(3))\ep^2}{4} \right )+\nonumber\\
+{\cal H}^3\left (2+\frac{\pi^2}{8}-\frac{(43+4\pi^2+32\zeta(3))\ep}{8} \right )+{\cal H}^4\left (\frac{15}{2}+\frac{7\pi^2}{16}+\frac{5\zeta(3)}{2} \right ) + \ldots
\end{eqnarray} 
Using the simple relation \eqref{gi} between the $\beta$-function and $\gamma_\nu$, from the condition $\beta({\cal H}_*)=0$, we obtain an expression for the value of the charge at a fixed point
\begin{eqnarray}
\label{fixed_point_ans}
{\cal H}_*=\frac{2\ep}{3}-\frac{2\ep^2}{9}+
\left (\frac{2}{9}-\frac{2\pi^2}{27} \right )\ep^3+\left (\frac{7}{81}+\frac{2\pi^2}{81}-\frac{16\zeta(3)}{81} \right )\ep^4 + \ldots
\end{eqnarray}
}
Finally, from \eqref{omega_def} we find the expression for the exponent $\omega$ in the fourth order of the $\varepsilon$-expansion
\begin{equation}
\label{omega_ans}
\omega=2\ep+\frac{2}{3}\ep^2+\frac{10}{9}\ep^3+\frac{56}{27}\ep^4 + \ldots
\end{equation}

In the considered stochastic turbulence model, the parameter $\varepsilon$ is not related to the space dimension $d$, and there is no critical dimension $d_c$, similar to that introduced in the theory of phase transitions. In this case, $d_c$ could be understood as the dimension at which Kolmogorov’s phenomenological theory becomes valid (anomalous scaling vanishes). There are a few arguments that indicate that this dimension is $d_c=\infty$, and therefore a number of works have shown interest in studying the asymptotic behavior of the stochastic turbulence model at $d\rightarrow\infty$ \cite{Kraichnan_1974,Frisch_Fournier_1978,Yakhot_1998, FRISCH1994, Runov_1999, AAKV03}. One may expect \cite{Frisch_Fournier_1978} that in the limit $d\rightarrow\infty$ the problem will be simplified and, possibly, will turn out to be exactly solvable, so that it can be used as a zero approximation of systematic perturbation theory with a small parameter $1/d$. 
However, a definite result has not been obtained yet. Significant progress in this regard has been achieved in the simplified Kazantsev-Kraichnan model of turbulent mixing of a passive admixture. In the work \cite{Kraichnan_1974}, it was proven that in this model the anomalous scaling disappears in the limit $d\rightarrow\infty$, and the corresponding anomalous exponents were calculated in the $1/d$ order. It was also possible to obtain these exponents within the framework of the $\epsilon$ expansion: in the first order it was done in \cite{AAV98} and then up to $\varepsilon^3$ in \cite{AABKV01}. It turned out that already in the first order $\varepsilon$-expansion correctly reproduces the result in the order $1/d$.

In the stochastic theory of turbulence, we are very interested in obtaining $1/d$-expansion, but for now calculations are possible only in double $(1/d,\ep)$-expansion, in leading order on $1/d$ and up to 3 order on $\ep$ \cite{AAGKK08, AAGKK09_turb_3l}.
One of the features of leading order on $1/d$ is 
notable decrease in the number of diagrams. 

Also, some interesting points were noted in the results obtained. For the critical exponent $\omega$, all irrational expressions present in the RG functions were cancelled out. In addition, a “reduction” of the diagrams was noticed; their significant mutual reduction results in the fact that in the two-loop approximation, the final result was determined by 4 diagrams out of 6, and in the three-loop approximation, by 9 diagrams out of 83 \cite{KK16}. To understand the mechanism of such reduction, it would be interesting to investigate its manifestation in higher orders of perturbation theory. However, the four-loop calculation was carried out only numerically \cite{Kirienko_pc}, which significantly narrows the possibilities of such an analysis.

As can be seen from \eqref{omega_ans}, the fourth term of the expansion of the exponent $\omega$ demonstrates the same property as the previous ones, namely, all irrational contributions in it have been cancelled out. Using the obtained analytical answers for the diagrams, we are planning to further analyze the reduction mechanism as well as perform a five-loop calculation.

The hyperlogarithm method and its implementation \textit{HyperInt} turned out to be very efficient in the model considered, allowing us to calculate all 4-loop diagrams. Due to the diagonalization of the determinant in the Feynaman representation in the asymptotics $d\rightarrow \infty$, the overwhelming number of diagrams in the next 5-loop approximation will highly likely turn out to be linear reducible, making it possible to use the same method for their calculation. Moreover, analytical calculations using HyperInt require  significantly less time compared to numerical calculations in this model.

Additionally, the results of this work give hope for carrying out analytical calculations in a similar way in other problems of critical dynamics, in which the currently highest achieved answers were mainly obtained by numerical methods (e.g. Sector Decomposition).

\section*{Acknowledgments}
We are deeply grateful to Yu. Kirienko for providing us his numeric four-loop result, which served as a foundation of this work as well as a valuable check of our calculations, and to E.~Panzer for his comments on hyperlogarithms and for providing the latest version of  \textit{HyperInt}. We would also like to thank E. Zerner-Käning for careful reading and editing. 

This work was performed at the Saint Petersburg Leonhard Euler International Mathematical Institute and supported by the Ministry of Science and Higher Education of the Russian Federation (agreement {
no.~075-15-2022-287}).


\section{Appendix}

The four-loop calculation of the exponent $\omega$ was also performed numerically~\cite{Kirienko_pc}. Like in the previous paper with the three-loop result \cite{AKKS17_turb}, the renormalization condition \eqref{oldtn} was used along with the function $F(m/k)=\theta(k-m)$ in \eqref{df2}.  The first three coefficients of the $\varepsilon$-expansion were obtained analytically, while the fourth one was calculated numerically. {
Written in terms of a charge $U=u/4$, where $u$ is the given charge of~\cite{Kirienko_pc},  the result for the anomalous dimension is}
\begin{eqnarray}
\label{Kir_res}
\gamma_\nu=U+\frac{U^2}{2}\left (1-2\ln2\cdot\ep+\frac{\pi^2}{6}\ep^2\right )+\nonumber\\+\frac{U^3}{8}\left (7+6\ln2+\ep\Big(8-\frac{\pi^2}{2}-45\ln3+24\ln2-18(\ln2)^2-9\;\mathrm{dilog}\frac{3}{4}\Big)\right ) +a_4U^4, \\  a_4=6.607378 \;. \quad\nonumber
\end{eqnarray} 
For the exponent $\omega$, this gives 
\begin{equation}
\omega=2\ep+\frac{2}{3}\ep^2+\frac{10}{9}\ep^3+2.0740695\ep^4+\ldots
\end{equation} 
The numeric coefficient of $\ep^4$ is in good agreement with the exact result obtained in this paper: $\frac{56}{27}=2.074074074...$. Given the exact value, one can reconstruct the analytical expression for the coefficient of $U^4$ in \eqref{Kir_res} by requiring the cancellation of the irrational terms:
\begin{equation}
a_4=\frac{9}{8}\Big(1+5\ln3-\frac{8}{3}\ln2+2(\ln2)^2+\mathrm{dilog}\frac{3}{4}\Big)\;.
\end{equation}

\bibliographystyle{JHEPsortdoi}

\bibliography{main}

\providecommand{\href}[2]{#2}\providecommand{\eprintlink}[2]{\href{#1}{#2}}\begingroup\begin{thebibliography}{10}

\bibitem{V04}
A.~N. Vasil’ev, {\emph{Quantum field renormalization group in critical behavior theory and stochastic dynamics}}.
\newblock Chapman and Hall/CRC, London, 2004.
\newblock originally published in Russian in 1998 by St.~Petersburg Institute of Nuclear Physics Press; translated by Patricia A.~de Forcrand-Millard.

\bibitem{Zinn_book}
J.~Zinn-Justin, {\emph{Quantum Field Theory and Critical Phenomena}}.
\newblock Clarendon Press, Oxford, 2002.

\bibitem{GGP_5l}
A.~Georgoudis, V.~Gonçalves, E.~Panzer, R.~Pereira, A.~V. Smirnov and V.~A. Smirnov, \href{\detokenize{http://dx.doi.org/https://doi.org/10.1007/JHEP09(2021)098}}{\textit{Glue-and-cut at five loops}},  \href{\detokenize{http://dx.doi.org/https://doi.org/10.1007/JHEP09(2021)098}}{\emph{J. High Energ. Phys.} \textbf{98} }(2021).

\bibitem{Lee5l}
R.~N. Lee, A.~V. Smirnov, V.~A. Smirnov and M.~Steinhauser, \href{\detokenize{https://api.semanticscholar.org/CorpusID:119241095}}{\textit{Four-loop quark form factor with quartic fundamental colour factor}},  \href{\detokenize{https://api.semanticscholar.org/CorpusID:119241095}}{\emph{J. High Energ. Phys.} \textbf{2019} (2019) }pp.~1--16.

\bibitem{Laporta2017}
S.~Laporta, \href{\detokenize{http://dx.doi.org/https://doi.org/10.1016/j.physletb.2017.06.056}}{\textit{High-precision calculation of the 4-loop contribution to the electron $g-2$ in {QED}}},  \href{\detokenize{http://dx.doi.org/https://doi.org/10.1016/j.physletb.2017.06.056}}{\emph{Phys.Lett.B} \textbf{772} (2017) }pp.~232--238.

\bibitem{KP17}
M.~V. Kompaniets and E.~Panzer, \href{\detokenize{http://dx.doi.org/10.1103/PhysRevD.96.036016}}{\textit{Minimally subtracted six-loop renormalization of {$O(n)$}-symmetric $\phi^4$ theory and critical exponents}},  \href{\detokenize{http://dx.doi.org/10.1103/PhysRevD.96.036016}}{\emph{Phys.Rev.D} \textbf{96} (2017) }p.~036016.

\bibitem{Schnetz7l}
O.~Schnetz, \href{\detokenize{http://dx.doi.org/10.1103/PhysRevD.107.036002}}{\textit{$\phi^4$ theory at seven loops}},  \href{\detokenize{http://dx.doi.org/10.1103/PhysRevD.107.036002}}{\emph{Phys.Rev.D} \textbf{107} (2023), no.~3 }p.~036002, \eprintlink{http://arxiv.org/abs/2212.03663}{arXiv:2212.03663 [hep-th]}.

\bibitem{Bednyakov2021}
A.~Bednyakov and A.~Pikelner, \href{\detokenize{http://dx.doi.org/10.1007/JHEP04(2021)233}}{\textit{Six-loop beta functions in general scalar theory}},  \href{\detokenize{http://dx.doi.org/10.1007/JHEP04(2021)233}}{\emph{JHEP} \textbf{04} (2021) }p.~233, \eprintlink{http://arxiv.org/abs/2102.12832}{arXiv:2102.12832 [hep-ph]}.

\bibitem{Bednyakov:2022}
A.~Bednyakov and A.~Pikelner, \href{\detokenize{http://dx.doi.org/10.1103/PhysRevD.106.076015}}{\textit{Six-loop anomalous dimension of the {$\phi^Q$} operator in the {$O(N)$} symmetric model}},  \href{\detokenize{http://dx.doi.org/10.1103/PhysRevD.106.076015}}{\emph{Phys.Rev.D} \textbf{106} (2022), no.~7 }p.~076015, \eprintlink{http://arxiv.org/abs/2208.04612}{arXiv:2208.04612 [hep-th]}.

\bibitem{Forster_Nelson77}
D.~Forster, D.~Nelson and M.~Stephen, \href{\detokenize{http://dx.doi.org/10.1103/PhysRevA.16.732}}{\textit{Large-distance and long-time properties of a randomly stirred fluid}},  \href{\detokenize{http://dx.doi.org/10.1103/PhysRevA.16.732}}{\emph{Phys.Rev.A} \textbf{16} (1977) }p.~732.

\bibitem{Folk_modelC}
R.~Folk and G.~Moser, \href{\detokenize{http://dx.doi.org/10.1103/PhysRevLett.91.030601}}{\textit{Critical dynamics of model {C} resolved}},  \href{\detokenize{http://dx.doi.org/10.1103/PhysRevLett.91.030601}}{\emph{Phys. Rev. Lett.} \textbf{91} (2003) }p.~030601.

\bibitem{Adzhemyan1999Hmodel}
L.~T. Adzhemyan, A.~N. Vasil'iev, Y.~S. Kabrits and M.~V. Kompaniets, \href{\detokenize{https://api.semanticscholar.org/CorpusID:120289875}}{\textit{{H}-model of critical dynamics: Two-loop calculations of {RG} functions and critical indices}},  \href{\detokenize{https://api.semanticscholar.org/CorpusID:120289875}}{\emph{Theor. Math. Phys.} \textbf{119} (1999) }pp.~454--470.

\bibitem{Dominicis_Hmodel}
C.~De~Dominicis and L.~Peliti, \href{\detokenize{https://journals.aps.org/prb/abstract/10.1103/PhysRevB.18.353}}{\textit{Field-theory renormalization and critical dynamics above {$T_c$}: Helium, antiferromagnets, and liquid-gas systems}},  \href{\detokenize{https://journals.aps.org/prb/abstract/10.1103/PhysRevB.18.353}}{\emph{Phys.Rev.B} \textbf{18} (1978) }p.~353.

\bibitem{AV84}
N.~V. Antonov and A.~N. Vasil’ev, \href{\detokenize{http://dx.doi.org/https://doi.org/10.1007/BF01018251}}{\textit{Critical dynamics as a field theory}},  \href{\detokenize{http://dx.doi.org/https://doi.org/10.1007/BF01018251}}{\emph{Theor.Math.Phys.} \textbf{60} (1984) }p.~671.

\bibitem{perc2023}
L.~T. Adzhemyan, M.~Hnatič, E.~V. Ivanova, M.~V. Kompaniets, T.~Lučivjanský and L.~Mižišin, \href{\detokenize{http://dx.doi.org/10.1103/PhysRevE.107.064138}}{\textit{Field-theoretic analysis of directed percolation: Three-loop approximation}},  \href{\detokenize{http://dx.doi.org/10.1103/PhysRevE.107.064138}}{\emph{Phys.Rev.E} \textbf{107} (2023) }p.~064138.

\bibitem{ANS08}
L.~T. Adzhemyan, S.~V. Novikov and L.~Sladkoff, {\textit{Calculation of dynamical exponent in model a of critical dynamics to order $\varepsilon^4$}},  {\emph{Vestn.St.Petersbg.Univ., Phys.Chem} (2008), no.~4 }pp.~110--114.

\bibitem{AEHIKKZ22}
L.~T. Adzhemyan, D.~A. Evdokimov, M.~Hnatič, E.~V. Ivanova, M.~V. Kompaniets, A.~Kudlis and D.~V. Zakharov, \href{\detokenize{http://dx.doi.org/https://doi.org/10.1016/j.physa.2022.127530}}{\textit{Model {A} of critical dynamics: 5-loop $\ep$ expansion study}},  \href{\detokenize{http://dx.doi.org/https://doi.org/10.1016/j.physa.2022.127530}}{\emph{Phys.A:Stat.Mech.Appl.} \textbf{600} (2022) }p.~127530.

\bibitem{AEHIKKZ22_2}
L.~T. Adzhemyan, D.~A. Evdokimov, M.~Hnatič, E.~V. Ivanova, M.~V. Kompaniets, A.~Kudlis and D.~V. Zakharov, \href{\detokenize{http://dx.doi.org/https://doi.org/10.1016/j.physleta.2021.127870}}{\textit{The dynamic critical exponent z for {2D} and {3D} ising models from five-loop $\ep$ expansion}},  \href{\detokenize{http://dx.doi.org/https://doi.org/10.1016/j.physleta.2021.127870}}{\emph{Phys.Lett.A} \textbf{425} (2022) }p.~127870.

\bibitem{DM79}
C.~De~Dominicis and P.~C. Martin, \href{\detokenize{https://link.aps.org/doi/10.1103/PhysRevA.19.419}}{\textit{Energy spectra of certain randomly-stirred fluids}},  \href{\detokenize{https://link.aps.org/doi/10.1103/PhysRevA.19.419}}{\emph{Phys.Rev.A} \textbf{19} (1979) }p.~419.

\bibitem{AAKV03}
L.~T. Adzhemyan, N.~V. Antonov, M.~V. Kompaniets and A.~N. Vasil'ev, \href{\detokenize{http://dx.doi.org/https://doi.org/10.1142/S0217979203018193}}{\textit{Renormalization-group approach to the stochastic {N}avier–{S}tokes equation: Two-loop approximation}},  \href{\detokenize{http://dx.doi.org/https://doi.org/10.1142/S0217979203018193}}{\emph{Int.J.Mod.Phys.B} \textbf{B17} (2003) }pp.~2137--2170.

\bibitem{AHKV05_2l}
L.~T. Adzhemyan, J.~Honkonen, M.~V. Kompaniets and A.~N. Vasil'ev, \href{\detokenize{http://dx.doi.org/10.1103/PhysRevE.71.036305}}{\textit{Improved $\ensuremath{\epsilon}$ expansion for three-dimensional turbulence: Two-loop renormalization near two dimensions}},  \href{\detokenize{http://dx.doi.org/10.1103/PhysRevE.71.036305}}{\emph{Phys.Rev.E} \textbf{71} (2005) }p.~036305.

\bibitem{AHKS05_Prandtl2l}
L.~T. Adzhemyan, J.~Honkonen, T.~L. Kim and L.~Sladkoff, \href{\detokenize{http://dx.doi.org/10.1103/PhysRevE.71.056311}}{\textit{Two-loop calculation of the turbulent {P}randtl number}},  \href{\detokenize{http://dx.doi.org/10.1103/PhysRevE.71.056311}}{\emph{Phys.Rev.E} \textbf{71} (2005) }p.~056311.

\bibitem{Prandtl2l_magn}
E.~Jurčišinová, M.~Jurčišin and R.~Remecký, \href{\detokenize{http://dx.doi.org/10.1103/PhysRevE.84.046311}}{\textit{Turbulent magnetic prandtl number in kinematic magnetohydrodynamic turbulence: Two-loop approximation}},  \href{\detokenize{http://dx.doi.org/10.1103/PhysRevE.84.046311}}{\emph{Phys.Rev.E} \textbf{84} (2011) }p.~046311.

\bibitem{AAGKK08}
L.~T. Adzhemyan, N.~V. Antonov, P.~B. Gol'din, T.~L. Kim and M.~V. Kompaniets, \href{\detokenize{http://dx.doi.org/10.1088/1751-8113/41/49/495002}}{\textit{Renormalization group in the infinite-dimensional turbulence: third-order results}},  \href{\detokenize{http://dx.doi.org/10.1088/1751-8113/41/49/495002}}{\emph{J. Phys. A: Math. Theor.} \textbf{41} (2008) }p.~495002.

\bibitem{Brown08}
F.~C.~S. Brown, \href{\detokenize{https://api.semanticscholar.org/CorpusID:15972133}}{\textit{The massless higher-loop two-point function}},  \href{\detokenize{https://api.semanticscholar.org/CorpusID:15972133}}{\emph{Commun. Math. Phys.} \textbf{287} (2008) }pp.~925--958.

\bibitem{Brown09}
F.~C.~S. Brown, \href{\detokenize{http://dx.doi.org/10.24033/asens.2099}}{\textit{Multiple zeta values and periods of moduli spaces $\overline{\mathfrak {m}}_{0,n}$}},  \href{\detokenize{http://dx.doi.org/10.24033/asens.2099}}{\emph{Ann. Sci. Ec. Norm. Super.} \textbf{Ser. 4, 42} (2009), no.~3 }pp.~371--489.

\bibitem{Brown09_arx}
F.~C.~S. Brown, \href{\detokenize{http://arxiv.org/abs/0910.0114}}{``On the periods of some feynman integrals.''} {}, \eprintlink{http://arxiv.org/abs/0910.0114}{arXiv:0910.0114 [math.AG]}.

\bibitem{Panzer14_main}
E.~Panzer, \href{\detokenize{http://dx.doi.org/https://doi.org/10.1016/j.cpc.2014.10.019}}{\textit{Algorithms for the symbolic integration of hyperlogarithms with applications to feynman integrals}},  \href{\detokenize{http://dx.doi.org/https://doi.org/10.1016/j.cpc.2014.10.019}}{\emph{Comput. Phys. Commun.} \textbf{188} (2015) }pp.~148--166.

\bibitem{Panzer_thesis}
E.~Panzer, \href{\detokenize{http://arxiv.org/abs/1506.07243}}{\emph{Feynman integrals and hyperlogarithms}}.
\newblock PhD thesis, Humboldt-Universität zu Berlin, 2015.
\newblock \eprintlink{http://arxiv.org/abs/1506.07243}{arXiv:1506.07243 [math-ph]}.

\bibitem{Schnetz23}
O.~Schnetz, \href{\detokenize{http://dx.doi.org/10.1103/PhysRevD.107.036002}}{\textit{{${\ensuremath{\phi}}^{4}$} theory at seven loops}},  \href{\detokenize{http://dx.doi.org/10.1103/PhysRevD.107.036002}}{\emph{Phys.Rev.D} \textbf{107} (2023) }p.~036002.

\bibitem{AMPS16}
A.~von Manteuffel, E.~Panzer and R.~M. Schabinger, \href{\detokenize{http://dx.doi.org/10.1103/PhysRevD.93.125014}}{\textit{Computation of form factors in massless {QCD} with finite master integrals}},  \href{\detokenize{http://dx.doi.org/10.1103/PhysRevD.93.125014}}{\emph{Phys.Rev.D} \textbf{93} (2016) }p.~125014.

\bibitem{AMPS20}
A.~von Manteuffel, E.~Panzer and R.~M. Schabinger, \href{\detokenize{http://dx.doi.org/10.1103/PhysRevLett.124.162001}}{\textit{Cusp and collinear anomalous dimensions in four-loop {QCD} from form factors}},  \href{\detokenize{http://dx.doi.org/10.1103/PhysRevLett.124.162001}}{\emph{Phys.Rev.Lett.} \textbf{124} (2020) }p.~162001.

\bibitem{AMPS21}
B.~Agarwal, A.~von Manteuffel, E.~Panzer and R.~M. Schabinger, \href{\detokenize{http://dx.doi.org/https://doi.org/10.1016/j.physletb.2021.136503}}{\textit{Four-loop collinear anomalous dimensions in {QCD} and {N}=4 super yang-mills}},  \href{\detokenize{http://dx.doi.org/https://doi.org/10.1016/j.physletb.2021.136503}}{\emph{Phys.Lett.B} \textbf{820} (2021) }p.~136503.

\bibitem{AKNS13}
L.~T. Adzhemyan, M.~V. Kompaniets, S.~V. Novikov and V.~K. Sazonov, \href{\detokenize{http://dx.doi.org/https://doi.org/10.1007/s11232-013-0057-6}}{\textit{Representation of the $\beta$-function and anomalous dimensions by nonsingular integrals: Proof of the main relation}},  \href{\detokenize{http://dx.doi.org/https://doi.org/10.1007/s11232-013-0057-6}}{\emph{Theor.Math.Phys.} \textbf{175} }(2013).

\bibitem{Brown_Kreimer13j}
F.~C.~S. Brown and D.~Kreimer, \href{\detokenize{http://dx.doi.org/10.1007/s11005-013-0625-6}}{\textit{{Angles, Scales and Parametric Renormalization}}},  \href{\detokenize{http://dx.doi.org/10.1007/s11005-013-0625-6}}{\emph{Lett. Math. Phys.} \textbf{103} (2013) }pp.~933--1007, \eprintlink{http://arxiv.org/abs/1112.1180}{arXiv:1112.1180 [hep-th]}.

\bibitem{Golz13_thesis}
M.~Golz, \href{\detokenize{https://www2.mathematik.hu-berlin.de/~kreimer/wp-content/uploads/BA_Golz_final.pdf}}{\textit{Evaluation techniques for feynman diagrams}},  \href{\detokenize{https://www2.mathematik.hu-berlin.de/~kreimer/wp-content/uploads/BA_Golz_final.pdf}}{\emph{bachelor thesis, Humboldt Universität zu Berlin} }(2013).

\bibitem{AAVBook}
L.~T. Adzhemyan, N.~V. Antonov and A.~N. Vasil’ev, {\emph{Field theoretic renormalization group in fully developed turbulence}}.
\newblock CRC Press, 1999.

\bibitem{AAGKK09_turb_3l}
L.~T. Adzhemyan, N.~V. Antonov, P.~Gol’din, T.~L. Kim and M.~V. Kompaniets, \href{\detokenize{http://dx.doi.org/10.1007/s11232-009-0032-4}}{\textit{Renormalization group in the theory of turbulence: Three-loop approximation as $d \to\infty$}},  \href{\detokenize{http://dx.doi.org/10.1007/s11232-009-0032-4}}{\emph{Theor. Math. Phys.} \textbf{158} (2009) }pp.~391--405.

\bibitem{AKKS17_turb}
L.~T. Adzhemyan, T.~L. Kim, M.~V. Kompaniets and V.~K. Sazonov, \href{\detokenize{http://dx.doi.org/https://doi.org/10.17586/2220-8054-2015-6-4-461-469}}{\textit{Renormalization group in the infinite-dimensional turbulence: determination of the {RG}-functions without renormalization constants}},  \href{\detokenize{http://dx.doi.org/https://doi.org/10.17586/2220-8054-2015-6-4-461-469}}{\emph{Nanosyst.: Phys. Chem. Math.} \textbf{6} (2015) }p.~461–469.

\bibitem{KP16}
M.~V. Kompaniets and E.~Panzer, \href{\detokenize{http://dx.doi.org/https://doi.org/10.22323/1.260.0038}}{\textit{Renormalization group functions of $\phi^4$ theory in the ms-scheme to six loops}},  \href{\detokenize{http://dx.doi.org/https://doi.org/10.22323/1.260.0038}}{\emph{PoS Proc. Sci. LL2016} }(2016).

\bibitem{AIKV_2018_4l}
L.~T. Adzhemyan, E.~V. Ivanova, M.~V. Kompaniets and S.~Y. Vorobyeva, \href{\detokenize{http://dx.doi.org/10.1088/1751-8121/aab20f}}{\textit{Diagram reduction in problem of critical dynamics of ferromagnets: 4-loop approximation}},  \href{\detokenize{http://dx.doi.org/10.1088/1751-8121/aab20f}}{\emph{J. Phys. A: Math. Theor.} \textbf{51} (2018), no.~15 }p.~155003.

\bibitem{Panzer14_diverg}
E.~Panzer, \href{\detokenize{http://dx.doi.org/10.1007/JHEP03(2014)071}}{\textit{{On hyperlogarithms and Feynman integrals with divergences and many scales}}},  \href{\detokenize{http://dx.doi.org/10.1007/JHEP03(2014)071}}{\emph{J. High Energy Phys.} \textbf{2014} (2014) }p.~71, \eprintlink{http://arxiv.org/abs/1401.4361}{arXiv:1401.4361 [hep-th]}.

\bibitem{chengwu}
H.~Cheng and T.~T. Wu, {\emph{Expanding Protons: Scattering at High Energies}}.
\newblock MA: MIT Press, Cambridge, 1987.

\bibitem{repo_turb}
\url{https://gitlab.com/Daniel999/turbulence}.

\bibitem{Kraichnan_1974}
R.~H. Kraichnan, \href{\detokenize{http://dx.doi.org/10.1017/S0022112074001881}}{\textit{Convection of a passive scalar by a quasi-uniform random straining field}},  \href{\detokenize{http://dx.doi.org/10.1017/S0022112074001881}}{\emph{J. Fluid Mech.} \textbf{64} (1974), no.~4 }p.~737–762.

\bibitem{Frisch_Fournier_1978}
J.~D. Fournier, U.~Frisch and H.~A. Rose, \href{\detokenize{http://dx.doi.org/10.1088/0305-4470/11/1/020}}{\textit{Infinite-dimensional turbulence}},  \href{\detokenize{http://dx.doi.org/10.1088/0305-4470/11/1/020}}{\emph{J. Phys. A: Math. Gen.} \textbf{11} (1978), no.~1 }p.~187.

\bibitem{Yakhot_1998}
V.~Yakhot, \href{\detokenize{http://dx.doi.org/10.48550/arXiv.chao-dyn/9805027}}{\textit{Strong turbulence in d-dimensions}},  \href{\detokenize{http://dx.doi.org/10.48550/arXiv.chao-dyn/9805027}}{\emph{arXiv:chao-dyn/9805027} }(1998).

\bibitem{FRISCH1994}
H.~L. Frisch and M.~Schulz, \href{\detokenize{http://dx.doi.org/https://doi.org/10.1016/0378-4371(94)90066-3}}{\textit{Turbulence effects in the high dimensionality limit}},  \href{\detokenize{http://dx.doi.org/https://doi.org/10.1016/0378-4371(94)90066-3}}{\emph{Phys. A: Stat. Mech. Appl.} \textbf{211} (1994), no.~1 }pp.~37--42.

\bibitem{Runov_1999}
A.~V. Runov, \href{\detokenize{http://dx.doi.org/10.48550/arXiv.chao-dyn/9906026}}{\textit{On the field theoretical approach to the anomalous scaling in turbulence}},  \href{\detokenize{http://dx.doi.org/10.48550/arXiv.chao-dyn/9906026}}{\emph{arXiv:chao-dyn/9805027} }(1999).

\bibitem{AAV98}
N.~V. Adzhemyan, L. Ts.~Antonov and A.~N. Vasil'ev, \href{\detokenize{http://dx.doi.org/10.1103/PhysRevE.58.1823}}{\textit{Renormalization group, operator product expansion, and anomalous scaling in a model of advected passive scalar}},  \href{\detokenize{http://dx.doi.org/10.1103/PhysRevE.58.1823}}{\emph{Phys.Rev.E} \textbf{58} (Aug, 1998) }pp.~1823--1835.

\bibitem{AABKV01}
L.~T. Adzhemyan, N.~V. Antonov, V.~A. Barinov, Y.~S. Kabrits and A.~N. Vasil'ev, \href{\detokenize{http://dx.doi.org/10.1103/PhysRevE.64.056306}}{\textit{Calculation of the anomalous exponents in the rapid-change model of passive scalar advection to order ${\ensuremath{\varepsilon}}^{3}$}},  \href{\detokenize{http://dx.doi.org/10.1103/PhysRevE.64.056306}}{\emph{Phys.Rev.E} \textbf{64} (2001) }p.~056306.

\bibitem{KK16}
Y.~V. Kirienko and T.~L. Kim, \href{\detokenize{http://dx.doi.org/10.21638/11701/spbu04.2016.202}}{\textit{The stochastic model of turbulence: Simplification of the diagram technique in high dimensions}},  \href{\detokenize{http://dx.doi.org/10.21638/11701/spbu04.2016.202}}{\emph{Vestn.St.Petersbg.Univ., Phys.Chem.} \textbf{3(61)} (2016) }pp.~151--157.

\bibitem{Kirienko_pc}
L.~T. Adzhemyan and Y.~Kirienko, \href{\detokenize{https://arxiv.org/abs/2406.14575}}{\textit{Theory of turbulence for $d\to\infty$: Four-loop approximation of the renormalization group}},  \href{\detokenize{https://arxiv.org/abs/2406.14575}}{\emph{arXiv:2406.14575} }(2024).

\end{thebibliography}\endgroup

\end{document}